\documentclass[a4paper,reprint,twocolumn,amsmath,amssymb,aps,floatfix]{revtex4}
\usepackage{graphicx}
\usepackage{color}

\newcommand{\ms}{\ensuremath{\mathrm{ms}}} 
\newcommand{\mm}{\ensuremath{\mathrm{mm}}} 
\newcommand{\cm}{\ensuremath{\mathrm{cm}}} 
\newcommand{\V}{\ensuremath{\mathrm{V}}}   
\newcommand{\E}{\vec E}                    
\newcommand{\n}{\vec n}                    
\newcommand{\en}{\hat{e}}                  
\newcommand{\fn}{\hat{f}}                  
\newcommand{\sn}{\hat{s}}                  
\newcommand{\Tn}{\hat{T}}                  
\newcommand{\VW}{\mathit{VW}}              
\newcommand{\VWn}{\overline{\VW}}          
\newcommand{\EG}{\mathit{EG}}              
\newcommand{\EGn}{\overline{\EG}}          
\renewcommand{\d}{\mathrm{d}}		   
\newcommand{\eq}[1]{\eqref{eq:#1}}         
\newcommand{\fig}[1]{Fig.~\ref{fig:#1}}    
\newcommand{\Fig}[1]{Figure~\ref{fig:#1}}  
\usepackage[normalem]{ulem}

\begin{document}
\title{Mechanisms of vortices termination in the cardiac muscle}
\author{%
  D. Hornung$^1$,
  V. N. Biktashev$^{2,6}$, %
  N. F. Otani$^{3}$, %
  T. K. Shajahan$^4$, %
  T. Baig$^1$, %
  S. Berg$^1$, %
  S. Han$^3$, %
  V. I. Krinsky$^{1,5,7}$, %
  S. Luther$^{1,8}$
}
\affiliation{ 
  $^1$Max Planck Institute DS, BMPG,  Germany. %
  $^2$University of Exeter, UK. %
  $^3$Rochester Institute of Technology, U.S.A. %
  $^4$National Institute of Technology Karnataka, India %
  $^5$INLN, CNRS, France.\\
  $^6$E-mail: V.N.Biktashev@exeter.ac.uk
  $^7$E-mail: Valentin.Krinsky@ds.mpg.de 
  $^8$E-mail: Stefan.Luther@ds.mpg.de
}   
\date{\today}
\begin{abstract}
We propose a solution to a long standing problem: how to
  terminate multiple vortices in the heart, when the locations of
  their cores and their critical time windows are unknown.  We
  scan the phases of all pinned vortices in parallel with electric
  field pulses (E-pulses). We specify a condition on pacing parameters
  that guarantees termination of one vortex. For more than one vortex
  with significantly different frequencies, the success of scanning
  depends on chance, and all vortices are terminated with a success
  rate of less than one. 
  We found that a similar mechanism terminates also a free (not
  pinned) vortex. A series of about 500 experiments with termination
  of ventricular fibrillation by E-pulses in pig isolated hearts
  is evidence that pinned vortices, hidden from direct
  observation, are significant in fibrillation. These results form a
  physical basis needed for the creation of new effective low energy
  defibrillation methods based on the termination of vortices underlying
  fibrillation. 
\end{abstract}
\maketitle

\section{Background}

Vortices play  crucial role in many domains of physics, including
catalytic waves, and condensed matter physics. In superconductors, the
motion of free vortices induces dissipation, so pinning is required to
maintain the superconductor state \cite{deGennesBook}. Pinning and
depinning transitions are essential features of superfluid dynamics
\cite{superfluid2000}.
 
Rotating electrical waves (vortices) and their instabilities underlie
cardiac chaos (fibrillation) \cite{%
  PertsovNature92,%
  WinfreeScience,%
  PertsovNature98%
}.  Physics of the vortices is well understood, e.g. \cite{%
  WinfreeNature84,%
  karma99,%
  KeenerTopolDef,%
  PanfilovPRL2008,%
  2012SteinbockPRL%
}. But the contemporary method of terminating the life-threatening cardiac
fibrillation is still aimed at termination of not vortices, but all
waves in the heart  \cite{HighVoltDefib94, HighVoltDefib97}.  It delivers a high energy electric shock, which
is damaging and painful.  Research aimed at reducing the energy to a
non-damaging, pain-free level gave rise to methods \cite{%
2004UnPinPRL,%
  pumir2007, %
  fenton2009,%
  luther2011,%
  Gray2011,%
  Janardhan2014%
} aimed at terminating vortices rather than all waves. We investigate
mechanisms of vortices termination by electric field pulses
(E-pulses). 

Over a century ago, it was found that a single vortex (rotating wave
or anatomical reentry) in a heart can be terminated with an electric
pulse \cite{mines1914}. An electrode was placed close to the
anatomical obstacle around which the wave rotates and a small energy
electric pulse was delivered within a certain time interval, called the
critical window, or vulnerable window, VW (note that for a rotating
wave, such intervals repeat within each lap). 

This approach alone can not terminate fibrillation since there are
multiple rotating waves with unknown and changing geometric locations
and phases \cite{PertsovNature98}.  That is, we have two main
problems: (i) the geometric positions of their cores, and (ii) the
positions of their critical time windows, are not known during
fibrillation.
An approach to overcome problem (i) 
has been previously developed \cite{%
  2004UnPinPRL,%
  pumir2007%
}.  Due to the bi-domain electric nature of cardiac muscle
\cite{sepulveda1989,  Sepulveda90}, every defect in it that can serve as a pinning
centre for a vortex, is at the same time an electric
inhomogeneity. This allows an E-pulse to excite the cores of all
pinned vortices simultaneously, regardless of the geometric positions
of their cores. 

\begin{figure}[b]
\includegraphics{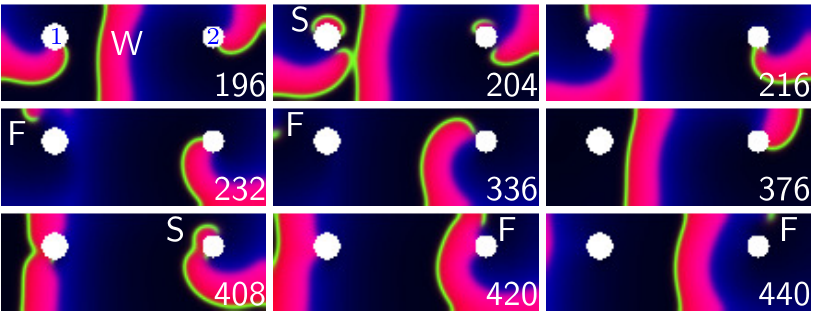}
\caption{%
  \textbf{Parallel termination of two pinned vortices with 
    geometrical locations and time positions of the critical
    (vulnerable) windows (VW) both unknown.} %
  The slow vortex 1 (period $T_{v1}=
  87\,\ms$, 
  pinned to the $1.2\,\mm$ defect 1, is entrained by the fast vortex 2
  (period $T_{v2}= 83\,\ms$, 
  $1.0\,\mm$ defect 2.  They are paced with electric field directed
  from top to bottom, $|\E|=1.3\,\V/\cm$, pulses $2\,\ms$ duration, period
  $100\,\ms$.  
  This induces the phase scanning with the time step
  $s=17\,\ms$. Color code: red is a wave, green is the wave
  front. Time is measured from the start of pacing at $t=0\,\ms$. 
  $\mathbf{196\,\ms}$: a wave W emitted by vortex 2 enslaves vortex
  1. %
  $\mathbf{204\,\ms}$: an E-pulse delivered at $t=200\,\ms$ induces a
  wave $S$. %
  $\mathbf{216}\,\ms$: the right wavebreak of wave $S$ annihilates
  with the tip of vortex 1 (they have opposite topological charges). %
  $\mathbf{232}\,\ms$: vortex 1 is unpinned and terminated. The left
  wavebreak of $S$ created a free vortex F. %
  $\mathbf{336\dots376\,\ms}$: F disappears on the boundary. %
  $\mathbf{408\dots440\,\ms}$: Next E-pulse similarly terminates
  vortex 2. %
  Barkley model, parameters $a=0.8$, $b=0.09$, $\epsilon=0.02$.
}
\label{fig:unknownVW}
\end{figure}

Approaches to resolve problem (ii) are being developed. They are
aimed to deliver a pulse into VWs of all vortices without knowing
their relative phases (by phases we mean the phases of
  oscillations, i.e. ``positions in time''). One of them is the
phase scanning by E-pulses, with a phase step that is shorter than the
VW, for all vortices in parallel. It was tested in experiment to
terminate a vortex in a rabbit heart preparation
\cite{Ripplinger2006}. Scanning with periodic E-pulses was used to
terminate fibrillation \cite{fenton2009,luther2011}. Termination of one
vortex with periodic E-pulses was numerically investigated in \cite{%
  2008Bittihn,%
  2010Bittihn,%
  behrend2010%
}.

\section{Theory}

\subsection{Pinned vortices}

In this paper, we investigate termination of multiple vortices
in the heterogeneous cardiac muscle.  The
difficulties arise due to the interaction of vortices.  We investigate
the excitation dynamics in the vicinity of the cores of pinned
vortices.  This allows us to draw conclusions about the overall dynamics.
When the VW of a vortex is hit by the E-pulse, this vortex is
displaced to a new position. If the vortex was situated close to the
tissue boundary, it is terminated.  Our aim is that VW of every vortex
is hit by an E-pulse (``all vortices are terminated'').
Wave patterns were calculated using the Barkley model
\begin{align*}
  u_t &=  {\varepsilon}^{-1}u(1-u)[u-(v+b)/a]+ \nabla^2 u,   \\
  v_t &= u-v 
\end{align*}
in a rectangular domain with circular holes, with no-flux boundary
conditions at the outer boundaries.
  
Pulses of electric field $\E$ are implemented as in \cite{pumir1999}
using the boundary conditions $\n \cdot(\nabla u - \E) = 0$ at the
boundaries of the holes.  The numerical integration used an explicit
Euler scheme with a time step of $1.6\cdot10^{-3}$ and
central-difference approximation of Laplacian with a space step of
$\frac{1}{6}$. The Barkley model is formulated in non-dimensional
units; for presentation purposes, we postulate that the time unit of
the Barkley model is $20\,\ms$ and the space unit of the Barkley model
is $0.5\,\mm$; this gives physiologically reasonable time and space
scales.
 
\Fig{unknownVW} shows termination of two pinned vortices by
E-pacing (see also the movie in the Supplementary Material).  This can
be achieved generically, for any parameters of the vortices, without
knowing their geometric location and time positions of the VWs.
    
\begin{figure}[tb]
\includegraphics{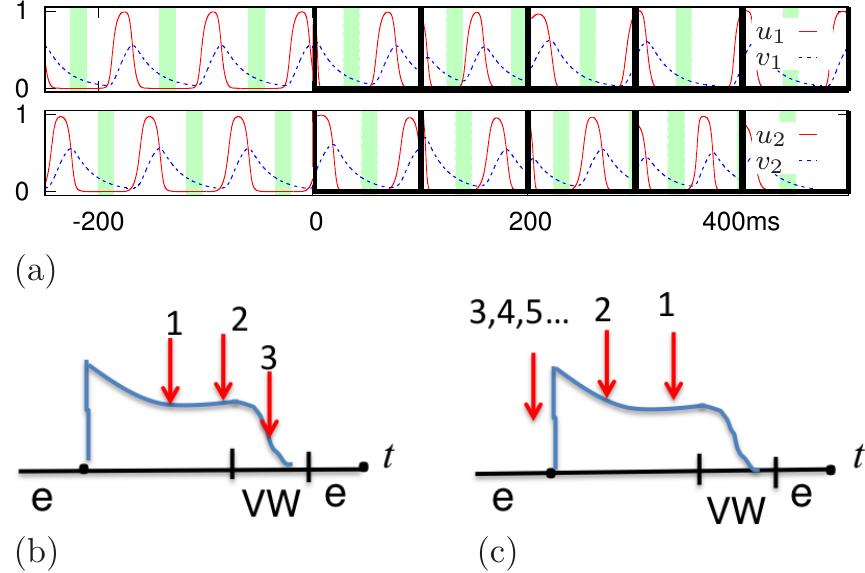}
\caption{%
  \textbf{Phase scanning}. %
  \textbf{(a)}: $u_1(t)$, $v_1(t)$ are recordings from the point just
  above defect 1, \fig{unknownVW}, and $u_2(t)$, $v_2(t)$ are
  same for defect 2.
  The bold black lines indicate timing of the delivered E-pulses.
  Shaded areas are vulnerable windows, defined as time intervals where
  $v\in(0.0871,0.18)$, $u<b/a$.
  Seen that in spite of small phase disturbances produced by E-pulses,
  the   topological features of the scanning are not disturbed, 
  scanning successfully terminates the vortices. E-pulse
  3 ($t=200\,\ms$) reaches VW of vortex 1 and terminates it, 
 compare with   \fig{unknownVW},
  $t=204\,\ms$, where is seen  a wave S induced by an E-pulse delivered at $t=200\,\ms$.
   E-pulse 5 
  ($t=400\,\ms$) reaches VW of vortex 2 and terminates it.  Compare
  with \fig{unknownVW}, 
  $t= 408\dots 440\,\ms$: E-pulse  delivered at  ($t=400\,\ms$) terminates vortex 2.
  \textbf{(b,c)}: schematic.  Superimposed action potentials (AP) are
  shown.
  Red arrows indicate timing of the delivered E-pulses; ``e'' is an
  excitable gap, $s$ is the scanning step, $s=T-T_v$. %
  \textbf{(b)} $s>0$ for $T >T_v$, scanning reaches the VW. %
  \textbf{(c)} $s<0$ for faster pacing $T<T_v$, the scanning moves in
  the opposite direction. E-pulse reaches the excitable gap ``e'',
  excites an AP thus resetting the rotation phase, and all subsequent
  pulses get into the same phase
  \cite{behrend2010,Shajahan-etal-2016}. 
  It does not reach the VW. %
}
\label{fig:scanning}
\end{figure}

To hit the VW with an E-pulse, the phase scanning
(\fig{scanning}a,b) should be performed with steps $0 <s<\VW$.
Thus, the VW duration (at the chosen $ \vec E $, see
\fig{edependence}h)
determines suitable values of $s$.  Then, the number of pulses $N$ 
required to
cover the whole period of a vortex is $N  \geqslant T_v/ s$, 
 where $T_v$ is the period of the vortex, $s=T-T_v$ is the scanning
step, and $T$ is the period of E-pacing. This gives the E-pacing
period $T=s+T_v$. Thus, all parameters of E-pacing ($E$, $N$, $T$) can
be set following equations
\begin{equation}
  0<s<\VW(E), 
  \quad
  N  \geqslant T_v/s,
  \quad
  T=s+T_v
  \label{eq:one}
\end{equation}
to guarantee that at least one E-pulse hits the VW.

Minimum energy for termination of  a pinned vortex is achieved when the
electric field strength is chosen so that the normalized vulnerable
window $\VWn(E)=1/N$, where $N$ is the number of pacing pulses.
The maximal success rate is achieved when the pacing frequency
$f=f_{best}$, where $f_{best}$ is the frequency for which the
normalized scanning step $\sn=1/N$.
When $f<f_{best}$, i.e. $T>T_{best} $, the scanning step $\sn>\VWn$,
and the vulnerable window may be missed while scanning, thereby decreasing
the success rate.  When $f>f_{best}$, so $T<T_{best}$, the scanning
step $\sn<\VWn=1/N$, and not all phases are scanned. This
also decreases the success rate.
  
What does interaction of vortices change here?  In cardiac muscle, the
fastest vortex entrains slower vortices if there is a normal wave
propagation between them.  Then, only one frequency remains; this
facilitates vortices termination.
But entrainment ceases if the fastest vortex is terminated before
the slower vortices, and then the frequency of the 
system changes (period increases). Here, two wave scenarios are possible, which we
describe for the case of just two vortices with periods $T_{v1}$ and
$T_{v2}$, such that $T_{v1}>T_{v2}$:
\begin{enumerate}
\item If the periods of the two vortices are not much different, so
  that $T_{v2}<T_{v1}<T$, then the pacing is still under-driving, and
  the slower vortex ($T_{v1}$) can be terminated by E-pacing with same
  period $T$ (see \fig{unknownVW} and
  \fig{scanning}b), provided that termination conditions
  \eq{one} are met for the slower vortex.
\item If however the periods of the two vortices are so much different that
\begin{equation}
                     T_{v2}<T<T_{v1}    \label{eq:failure}
\end{equation}
then the pacing with the same period is no longer under-driving, but
over-driving. And overdrive pacing will typically entrain the
remaining vortex rather than eliminate it.
\end{enumerate}
For successful termination of fibrillation, the E-pacing period should
be increased to a higher value $T_2$,  such that $T_{v1}<T_2$.  Thus,
vortices can be terminated in any case.  Experiments \cite{luther2011}
underestimated the potential of the method since this mechanism was
not known yet.
 
\subsection{Free vortices}

\begin{figure}[h]
\includegraphics{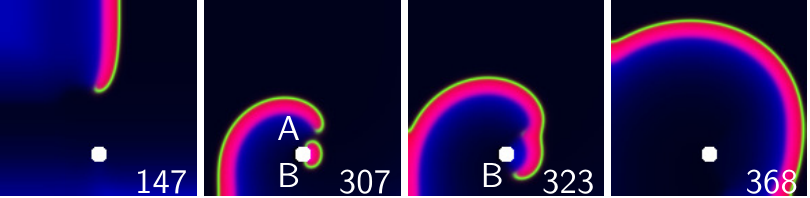}
\caption{%
  \textbf{Termination of a free vortex by an E-pulse.} %
  $\mathbf{147\,\ms}$: A free vortex and a defect (white). %
  $\mathbf{307\,\ms}$: A semi-circular wave (with wavebreaks A and
  B) emitted from the defect by an E-pulse, electric field directed
  from right to left. %
  $\mathbf{323\,\ms}$: Wavebreak A fused with the vortex tip. %
  $\mathbf{368\,\ms}$: After annihilation of wavebreak B with the
  border, only a wave without wavebreaks is left in the medium. %
  Barkley model, parameters $a=0.6$, $b=0.075$, $\epsilon=0.02$. %
}
\label{fig:freeSpiral}
\end{figure}

Below, we describe a mechanism terminating a high frequency
  free vortex by electric field pacing. Numerical and theoretical
  publications state it is very easy: usual local pacing (ATP) with
  frequency higher than a free vortex frequency terminates a free
  vortex. It works in experiment and in clinics, but only for low
  frequency vortices. Classic ATP cannot terminate VF, cannot
  terminate high frequency rotating waves, including free rotating
  waves.  Waves emitted from a pacing electrode propagate along the
  whole tissue only for low frequency.  For higher frequencies, the
  Wenckebach rhythm transformation arrives in a heterogeneous cardiac
  tissue. In contrast, an electric field penetrates everywhere,
  without frequency limitations,  and only requires local
    heterogeneities to act as virtual electrodes. This mechanism can
    be used for terminating a high
  frequency free vortex.
A  free (not pinned) vortex  can be terminated when
its moving core passes not very far (at distance $L \lesssim \lambda$,
where $\lambda$ is the wave length) from a defect in the medium,
serving as a virtual electrode, \fig{freeSpiral}
(this illustration uses the same mathematical model for the
  excitable medium and for the action of the electrical field as in
  the previous subsection).
The success rate increases as distance $L$
  decreases.
A mechanism reliably terminating a free rotating wave was found in
1983 \cite{Krinsky1983}: waves with a frequency higher than the
frequency of a rotating wave, induce its drift and termination on the
boundary. Cardiologists used a high-frequency pacing (anti-tachycardia
pacing, ATP) well before the mechanism was understood.  
But ATP can 
not terminate high frequency rotating waves.
This fundamental limitation is overcome by the mechanism of
  a free vortex termination proposed here. 
This mechanism depends on wave emission induced 
from a defect induced by the electric field. 
The electric field penetrates
everywhere, hence no restriction on its efficacy imposed by the maximal frequency
of propagating waves in any part of the cardiac tissue.
 
   An increased amplitude of
  electric field $|\E|$ results in defibrillation. There is a
  classical explanation: electric field should be increased to the
  value where it terminates all propagating waves.  A physical
  explanation is: the wave emission is induced from a larger number
  of defects when the electric field is increased \cite{pumir2007}.
  \Fig{edependence}g shows another mechanism: 
  the size of the excited region increases with the electric
    field, and the duration of the VW increases with it.

\subsection{Time separation analysis of VW}

The mechanism of the VW is
related to the change of topological charge in 1D and
the creation of new
topological singularities in 2D.

\begin{figure}[htbp]
\includegraphics{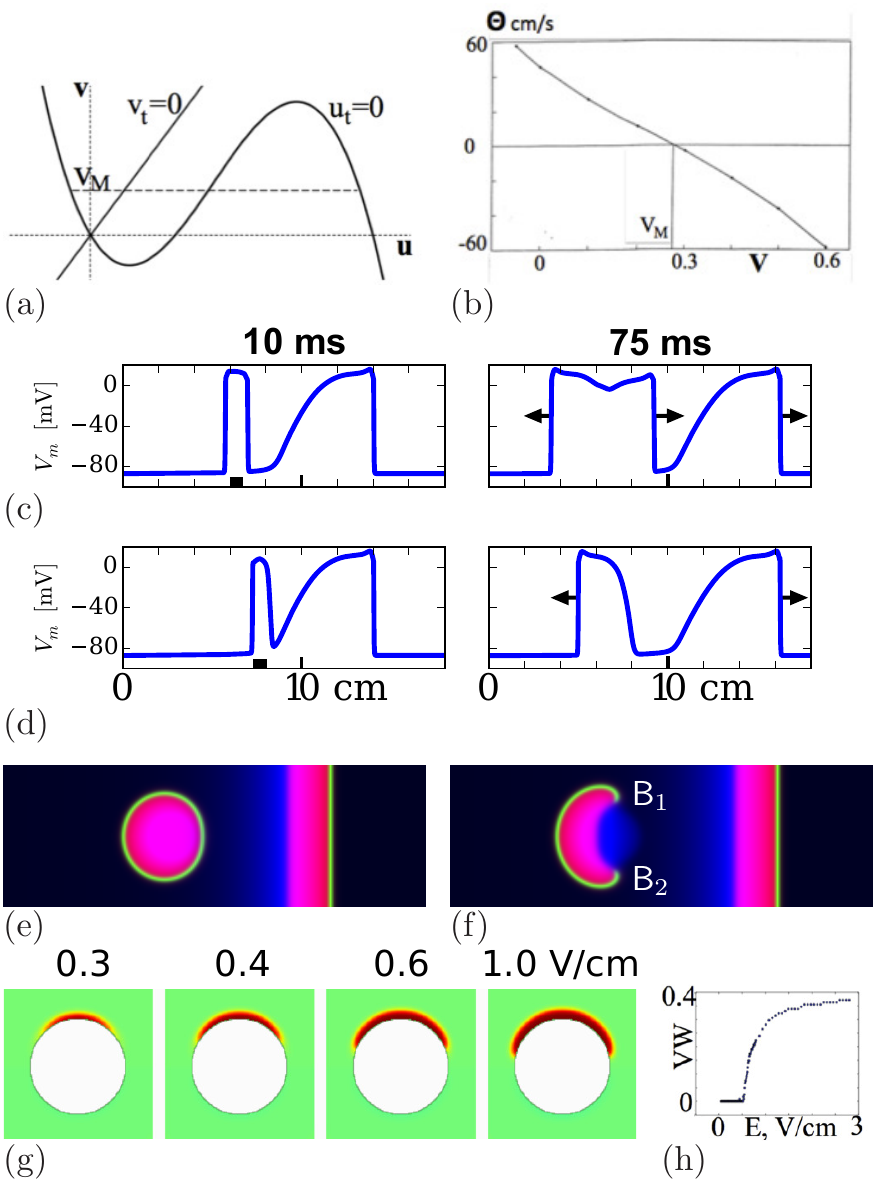}
\caption{ %
  \textbf{Change of topological charge, creation of phase
    singularities, and vulnerable window VW}. %
  \textbf{(a-d)} \textit{1 dim mechanism}. %
  \textbf{(a)} Nullclines of FHN equations. $M$ is the Maxwell point.
  The topological charge of a wave pattern is changed by an E-pulse
  only when an image of a nucleated wave contains the Maxwell point.
  \textbf{(b)} Wave front velocity $\theta$ vs the slow variable
  $v$. The value of $v$ corresponding to velocity $\theta=0$ is the
  ordinate $v_M$ of the Maxwell point on panel (a). %
  \textbf{(c)} The topological charge conservation in 1 dim. Generic
  case: an electric pulse $3\,\ms$ in duration is delivered far from the
  tail of an action potential (AP). %
  $t=10\,\ms$ after pacing: a nucleated wave, very narrow, and the
  electrode (black square) below it. %
  $t=75\,\ms$: the nucleated wave developed into two counter
  propagating APs. Their total topological charge is zero. %
  \textbf{(d)} Violation of the topological charge conservation. %
  $t=10\,\ms$: an electric pulse is delivered closer to the tail of
  the AP, inside VW. %
  $t=75\,\ms$: only one AP is induced. It propagates to the left
  only. The topological charge is changed. %
  Cardiac ionic model by Majahan et al. 2008 \cite{IonicModel}. %
  \textbf{(e,f)} {\it 2D mechanism}. %
  \textbf{(e)} No phase singularities are created. An electric pulse is
  delivered as in (c). %
  \textbf{(f)} Creation of two phase singularities, B1 and B2.  An
  electric pulse is delivered as in (d). %
  \textbf{(g,h)} VW increases with electric field in 2D. %
  \textbf{(g)} Mechanism: the larger $E$, the larger is the depolarised
  region. %
  \textbf{(h)} Graph $\VW(E)$. %
}
\label{fig:edependence}
\end{figure}
 
The topological charge 
    (or ``winding number'', in terminology of~\cite{KeenerTopolDef})
    in a 1D closed circuit
    can be defined as the increment of the excitation phase per one
  loop around the circuit, where the excitation phase can be defined
  as the angle in the polar coordinates in the phase plane of the
  reaction kinetics, centered at a suitably chosen point in the ``No
  Mans Land'', in terminology of~\cite{FitzHugh-1961}. Likewise,
  topological singularity in 2D is defined as a point such that any
  sufficiently small 1D contour surrounding it has a topological
  charge. We illustrate the relationship between these concepts and
  VW  using time-separation analysis for the FitzHugh-Nagumo (FHN)
equations:
\begin{eqnarray}
 u_t &=& f(u) - v + D u_{xx},  \label{eq:four}\\
 v_t &=& \varepsilon(u - kv) . \label{eq:five}
\end{eqnarray}
Here $f(u)= A u(1-u)(u-\alpha)$, and $\epsilon \ll 1$ is a small
parameter permitting the time-scales separation (for details of
relevant formalisms see review~\cite{tyson1988}). The wavefront
propagation velocity $\theta$ can be estimated by assuming that the
slow variable $v$ is approximately constant across the wavefront. The
propagation of the front is then described by Eq.~\eqref{eq:four}
alone, where $v$ is a constant parameter. Transforming the independent
variables such that $\xi=x-\theta t $ makes Eq.~\eqref{eq:four} an
ordinary differential equation
\[
 -\theta u_\xi = f(u) - v + D u_{\xi\xi}
\]
which together with boundary conditions $u(\infty)=u_1$,
$u(-\infty)=u_3$, where $u_1=u_1(v)$ and $u_3=u_3(v)$ are respectively
the lowest and highest roots of $f(u)=v$, define $\theta$ as a
function of $v$, see \fig{edependence}b.  Here, velocity
$\theta(v)$ is negative for $v>v_M$, where $v_M$ is the
Maxwell point, 
$\int_{u_1(v_M)}^{u_3(v_M)}\left(f(u)-v_M\right)\,\d{u}=0$,
$\theta(v_M)=0$~\cite{2004UnPinPRL}.

Vulnerability is a cardiological term coined for initiation of
fibrillation by an electric pulse.  In the physical language,
vulnerability in 1 dim can be related to a change of the topological
charge, and in 2D and 3D to creation of new phase singularites.  In
1 dim, this phenomenon happens when the current injection nucleates a
wave propagating in only one direction,
\fig{edependence}d. This is in contrast to the generic case,
where the topological charge is conserved, when the new wave
propagates in two directions, \fig{edependence}c, or new wave
is not nucleated at all (not shown).  For one-directional propagation
to happen, the nucleated wave should cover the points 
which have $v=v_M$ corresponding to the Maxwell
point $\theta=0$.  Then, a part of the nucleated wave has positive
velocity (becoming the front of the wave) and another part has a
negative velocity (becoming the tail of the wave), as in
\fig{edependence}d,f. Otherwise, all parts of the nucleated
wave have velocity of the same sign. When velocity $\theta <0$, the
nucleated wave shrinks and decays. In the opposite case, it enlarges
in all directions, as in \fig{edependence}c,e.

\subsection{Axiomatic model based on properties of VW}

 Now we formulate an axiomatic model based on the
properties of VW discussed above. 
  Let $\phi^j_n\in[0,1)$, $j=1,2$, $n=1,\dots,5$ describe the phase of
$j$-th vortex just after the delivery of the $n$-th E-pulse, $T_j$ be
the natural periods of the vortices, $T_2>T_1$, and correspondingly
$\sn_j=\Tn-\Tn_j=(T-T_j)/T_d$ are the scanning steps normalized by the
measured dominant period, $T_d$.  We postulate
$\phi^j_{n+1}=(\phi^j_n+s)\mod 1$, subject to the following
corrections: (i) if $\phi^j_{n+1}\in[1-\EGn,1)$, where $\EGn$ is the
normalized duration of the excitable gap, then $\phi^j_{n+1}$ is
replaced with 0: this describes resetting the $j$-th phase by the
E-pulse; (ii) if $\phi^j_{n+1}\in[1-\EGn-\VWn,1-\EGn)$, where $\VWn$
is the normalized duration of the vulnerable window, then the $j$-th
vortex is considered terminated; (iii) if neither vortex is
terminated, then the slower vortice's phase is enslaved by the faster
one's, $\phi^2_{n+1}=(\phi^1_{n+1}-D) \mod 1$, where $D$ is a fixed
phase delay; (iv) if both vortices are terminated, iterations stop and
E-pacing is deemed successful.

\begin{figure}[htb]
\includegraphics{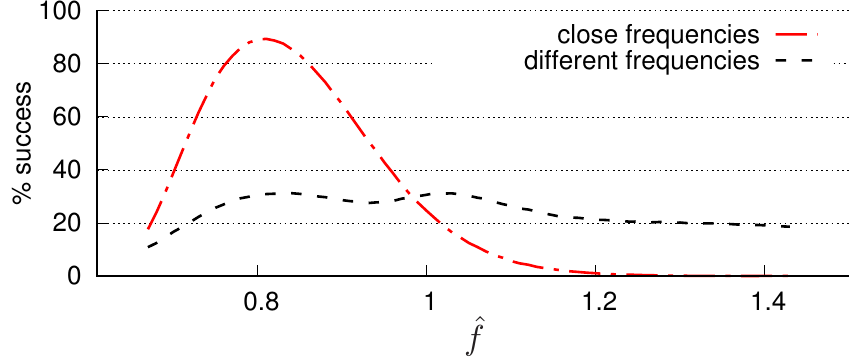}
\caption{%
  \textbf{Success rate of  two vortices termination.  }
    Success rate vs normalized frequency $\fn=f/f_d$ where $f_d$ is
    the dominant frequency.  On the image, ``different''
    and ``close'' frequencies mean significant and insignificant
    difference in frequencies as per Eq.\eq{failure}.  Numerical
    calculations with the normalized vulnerable window $\VWn$ =0.2.  }
\label{fig:theor6b}
\end{figure}

\Fig{theor6b} shows the success rate of termination
    of two vortices as a function of the normalized frequency of
  E-pacing. The graphs represent results of Monte-Carlo
  simulations of the axiomatic model described above, with
  random initial phases of vortices and  two variants for the
    choice of frequencies: (i) normal distributions of parameters
    $\Tn_1=1\pm0.1$ and $\Tn_2=1.6\pm0.05$ (mean$\pm$standard
    deviation), ``different frequencies'', and (ii) same parameters
    for $\Tn_1$, and $\Tn_2$ enforced very close to $\Tn_1$, namely
    $\Tn_2=(1+10^{-6})\Tn_1$, ``close frequencies'', with other
  parameters fixed at $\EGn=0.4$, $\VWn=0.2$, $D=0.25$.  
  The success rate of termination of two vortices
  with significant difference in frequencies as per Eq.\eq{failure} is
  seen in \fig{theor6b} to be three-fold lower than that
  for vortices with insignificant difference in
  frequencies. This happens because when the leading (fastest) vortex
  is terminated first, the same E-pacing period $T$ appears below the
  period $T_1$ of the resting slower vortex, see \eq{failure}.  Thus
  the resting vortex cannot be terminated, see \fig{scanning}c.

Vortices termination can be induced also by other mechanisms different
from vulnerability, e.g. pacing-induced drift of a free vortex
\cite{Krinsky1983}, unpinning of weakly pinned vortices \cite{%
  Marcel2008,%
  Pumir-krinsky2010%
} and by 3D mechanisms \cite{Biktashev-1989,Zemlin-2003,Otani-2016}.

\section{Experiment}

\begin{figure}[h]
\includegraphics{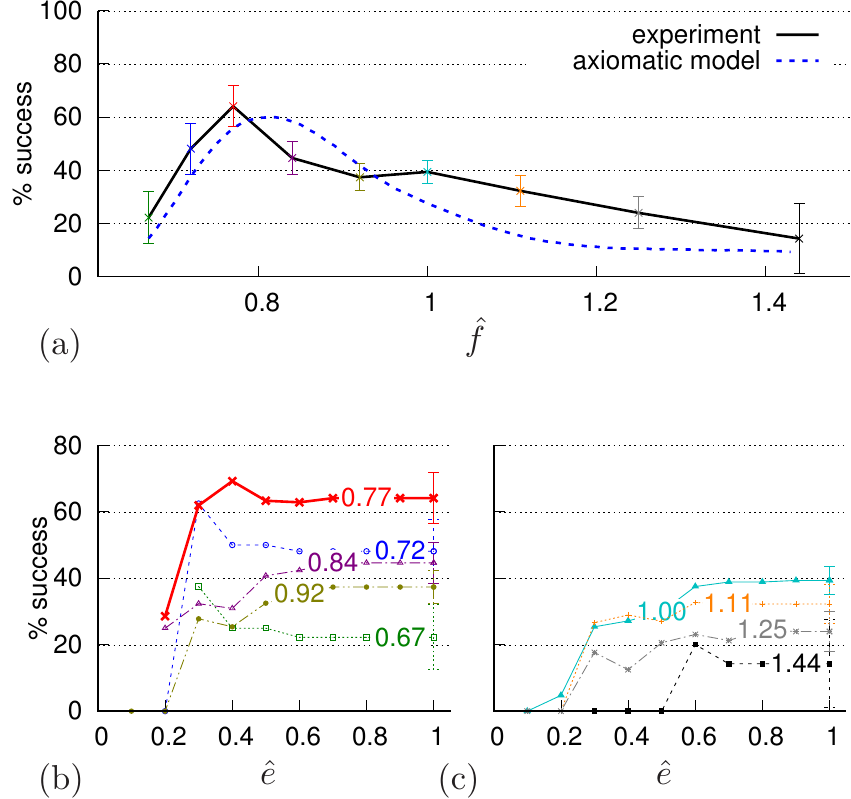}
\caption{%
  \textbf{Fibrillation termination in the isolated pig hearts}.  The
  success rate of defibrillation in 486 experiments by 5 biphasic
  E-pulses.  %
  \textbf{(a)} Success rate vs normalized frequency $\fn=f/f_d$ where
  $f_d$ is the dominant frequency of fibrillation. Error bars: the
  standard deviation.  %
  Blue  curve is obtained by mixture (50:50) of two theoretical curves shown in \fig{theor6b}.
     \textbf{(b,c)} Success rate for defibrillation energies not
  exceeding $\en$, for frequencies $\fn$ shown near each
  curve. Normalized energy $\en=e/e_1$, where $e_1$ is the threshold
  $E_{50}$ energy of defibrillation by 1 shock. 
  (b) shows graphs for   $\fn <1$, (c) shows graphs for   $\fn  \geqslant1$. 
  Graphs (b, c) and the experimental curve in (a) are calculated from data in  \cite{Hornung2013}.
  Image (a) indicates that in about a half of fibrillation
  experiments, the frequencies of the vortices were not significantly
  different.
  The optimal pacing frequency $\fn=0.77$ is below the arrhythmia
  frequency $(\fn <1)$ as it should be for terminating pinned
  vortices.  These experiments provide evidence that pinned vortices, hidden
  from direct observation, are significant in fibrillation.  }
\label{fig:pigExper}
\end{figure}

Results of about 500 experiments with vortices termination in the
isolated pig hearts are presented in \fig{pigExper}.
Fibrillation was induced and terminated as in \cite{%
  fenton2009,%
  luther2011%
}. %
In terms of the normalized pacing frequency
  $\fn$, the numbers $n$ of the experiments were: 
  $n=18$ for $\fn=0.67$;
  $n=28$ for $\fn=0.72$;
  $n=39$ for $\fn=0.77$;
  $n=65$ for $\fn=0.84$;
  $n=91$ for $\fn=0.92$;
  $n=127$ for $\fn=1$;
  $n=62$ for $\fn=1.11$,
  $n=50$ for $\fn=1.25$ and
  $n=7$ for $\fn=1.44$.

In \fig{pigExper}a the experimental curve is fit by the blue
theoretical curve much better than by either of the two theoretical
curves in \fig{theor6b}.  It indicates that in about a half of
fibrillation experiments, the frequencies of the vortices were not
significantly different.
    
\Fig{pigExper}a,b show that the optimal pacing frequency $\fn=0.77$ is
below the arrhythmia frequency $(\fn <1)$ as it should be for
terminating pinned vortices.  Notice that elimination of a free,
rather than pinned, vortex by inducing its drift via the mechanism
described in \cite{Krinsky1983}, requires the pacing frequency to be
above the arrhythmia frequency, $\fn >1$.  These experiments provide evidence
that pinned vortices, hidden from direct observation, are significant
in fibrillation. In particular, they show that the VW mechanism
is an explanation for the high success rate of VF termination using
electric field pacing.

\section{Discussion}
 
  In this paper, after more than 25 years of research, we
   propose a solution to a problem, how to terminate multiple vortices
   in the cardiac tissue hidden from direct observation.  In order to
   control vortices, two problems should be overcome: both, the
   geometric positions of their cores, and the positions of their
   critical time windows, are not known during fibrillation.  The
   first problem we have solved previously using an electric field
   pulse to excite the cores of all pinned vortices simultaneously.
   Approaches to solve the second problem are being developed.  One of
   them is based on the phase scanning of all pinned vortices in
   parallel to hit the critical time window of every pinned vortex.
   In this paper, we investigate the related physical mechanisms using
   simple two variable models as well as a detailed ionic model of the
   cardiac tissue.  A similar mechanism terminates also a free (not
   pinned) vortex, when the vortex's core passes not very far from a
   defect.

Even though it is widely believed that the
success of defibrillation has a probabilistic nature, we have shown
that termination of one vortex can be achieved deterministically, in any
case. This can be achieved generically, for any parameters of the
vortex, without knowing its geometric location and timing of its
VW.  All that is needed is to set the parameters of E-pacing ($E$,
$N$, $T$) according to equations~\eq{one}.
Termination of an arrhythmia becomes probabilistic
when two or more vortices are involved.  If there is normal wave
propagation between the two vortices, and the slower vortex is
enslaved by the faster one, 
then the E-pacing protocol described in \cite{luther2011}
cannot control which of the
vortices will be terminated first.
If the slower vortex is terminated first, the frequency of
the system does not change, and both vortices are terminated
deterministically, in any case. If, by chance, 
the faster vortex is terminated first, the frequency of the system 
changes, and the remaining slower vortex may be not
terminated if conditions~\eq{failure} are
satisfied.   

Here, we investigated two extreme cases: permanently
pinned vortices and permanently free vortices.  There is no sharp
transition between them  
in heterogeneous media with different size pinning centers.
In cardiac muscle, there are heterogeneities of all sizes, 
including those to which vortices pin weakly.  A weakly pinned vortex 
is pinned for some time only, then leaves the pinning center and moves 
as a free vortex, again for some time. When moving and meeting a 
pinning center, it may pin to it, or may reach the boundary of the 
tissue and disappear.  

3D models are widely used in investigation of wave patterns induced
by rotating waves, e.g.  \cite{2012SteinbockPRL}.  
A 3D mechanism of defibrillation was described in
  \cite{Biktashev-1989,Zemlin-2003,Otani-2016}. %
  Study of vortices
termination in 2D models is a necessary step for developing
understanding mechanisms of 3D vortices termination in the heart.
Termination vortices underlying fibrillation is only a small part of a
problem preventing and curing the cardiac arrhythmias where a
combination of molecular and dynamics approaches is prominent
\cite{Weiss-etal-2015-JMCC}.
 
In conclusion, we have shown mechanisms of terminating pinned and free
vortices by electric field pulses when the geometric positions of
their cores, and the phases of rotation are not known.  These results
form the physical basis for creation of new effective methods for
terminating vortices underlying fibrillation.

\textbf{Ethics.} %
The study was reviewed and approved by the ethics committee, permit
no. 33.9-42052-04-11/0384, Lower Saxony State Office for Customer
Protection and Food Safety.


\textbf{Data Accessibility.} %
A movie version of Figure 1 is available as Electronic Supplementary
Material.

\textbf{Competing interests.} %
We have no competing interests.

\textbf{Authors' contributions.} %
D.H. designed the experimental study of VF on pig hearts, carried out
the experiment and the data analysis. %
V.N.B. designed the theoretical and numerical study, carried out
analysis of the experiment data, helped draft and edit the
manuscript. 
N.F.O. designed the numerical study. %
T.K.S. designed the cell culture experiment, and performed the experiment. %
T.B. carried out experiment and the data analysis of VF termination on
pig heart and performed numerical simulations. %
S.B. designed and performed the numerical study of an ionic cardiac
model. %
S.H. ran the free vortex computer simulation. %
V.K. conceived of the theoretical study and wrote the manuscript. %
S.L. designed the experimental study, coordinated the whole study
and designed the experiment. %
All authors gave final approval for publication.

\textbf{Acknowledgements}
  We are grateful to R. F. Gilmour, Jr.,
  A. M. Pertsov and J. Wikswo  for constructive criticism and inspiring discussions.

\textbf{Funding.} %
The research leading to the results has received funding from Max
Planck Gesellschaft, the European Community Seventh Framework
Programme FP7/2007-2013 under Grant Agreement 17
No. HEALTH-F2-2009-241526, EUTrigTreat %
(DH, TB, SB, VIK, SL), %
and from EPSRC (UK) grant EP/I029664 %
(VNB).%
We also acknowledge support from the German Federal Ministry of
Education and Research (BMBF) (project FKZ 031A147, GO-Bio), the
German Research Foundation (DFG) (Collaborative Research Centres SFB
1002 Project C3 and SFB 937 Project A18), the German Center for
Cardiovascular Research (DZHK e.V.)
(DH, TB, SB, VIK, SL), %
Science \& Engineering Research Board of Department of Science \&
Technology, Government of India (TKS), %
EPSRC (UK) grant EP/N014391 (VNB) %
and U.S. NIH Grant No. R01HL089271 (NFO). %

\end{document}